\newcommand{\captionstyle}{\normalfont} 
\newcommand{\width}{0.5}

\documentclass[pra,aps,twocolumn,superscriptaddress]{revtex4-1}%

\usepackage{graphicx}
\usepackage{bm}
\usepackage{amsmath}
\usepackage{amssymb}
\usepackage{verbatim}
\usepackage[colorlinks=true, pdfstartview=FitV, linkcolor=blue, citecolor=blue, urlcolor=blue]{hyperref} 
\usepackage[english]{babel}
\usepackage{natbib}

\newcommand {\INLN} {Institut Non Lin\'{e}aire de Nice, CNRS and
Universit\'e Nice Sophia-Antipolis,\\ 1361 route des Lucioles, 06560
Valbonne, France}
\newcommand {\Brazil} {Departamento de F\'isica, Universidade Federal da Para\'iba - Caixa Postal 5086, 58051-900 Jo\~ao Pessoa-PB, Brazil}
\newcommand {\BBrazil} {Unidade Acad\^emica de F\'isica, Universidade Federal de Campina Grande, 58429-900 Campina Grande, PB, Brazil}
\newcommand {\LPMC} {Universit\'e Nice Sophia Antipolis, Laboratoire de Physique de la Mati\`ere Condens\'ee, CNRS UMR 7336, Parc Valrose, 06108 Nice, France}
\newcommand {\OCA} {Laboratoire Lagrange, UMR 7293, Universit\'e de Nice-Sophia Antipolis, CNRS, Observatoire de la C\^ote d'Azur, 06300 Nice, France}

\begin{document}


\title{Temporal intensity correlation of light scattered by a hot atomic vapor}

\author{A. Dussaux}
\email{antoine.dussaux@inln.cnrs.fr} 
\affiliation{\INLN}
\author{T. Passerat de Silans}
\affiliation{\Brazil}
\affiliation{\BBrazil}
\author{ W. Guerin}
\affiliation{\INLN}
\author{O. Alibart}
\affiliation{\LPMC}
\author{S. Tanzilli}
\affiliation{\LPMC}
\author{F. Vakili}
\affiliation{\OCA}
\author{R. Kaiser}
\affiliation{\INLN}

\date{\today}

\begin{abstract}
We present temporal intensity correlation measurements of light scattered by a hot atomic vapor. Clear evidence of photon bunching is shown at very short time-scales (nanoseconds) imposed by the Doppler broadening of the hot vapor. Moreover, we demonstrate that relevant information about the scattering process, such as the ratio of single to multiple scattering, can be deduced from the measured intensity correlation function. These measurements confirm the interest of temporal intensity correlation to access non-trivial spectral features, with potential applications in astrophysics. 

\end{abstract}

\maketitle
\section{Introduction}

Spatial intensity correlation measurements were first developed in astrophysics, where the correlation of light collected by two telescopes at variable remote locations made it possible to infer stellar angular diameters \cite{HBT56,HBT68}. Temporal intensity correlation measurements, or intensity correlation spectroscopy, based on a single telescope, from astrophysical light sources, has not been demonstrated so far due to technical challenges in terms of time resolution and spectral filtering, but may be useful for resolving narrow spectral features \cite{Johansson2005361,1301176}. On the other hand, temporal intensity correlation spectroscopy is widely exploited in quantum optics for the characterization of non-classical states of light \cite{LPOR04}, and is also of interest in cold-atom physics \cite{Jurczak,PhysRevA.53.3469,PhysRevA.68.013411,Stites:04,PhysRevA.92.013850,Nakayama:10}. Ideal photon bunching for classical light, i.e., with maximum temporal contrast, has even recently been reported in a cold-atom experiment \cite{Nakayama:10}. One of the main challenge in intensity correlation measurements lies in time resolution of the detection, which needs to be on the order of the coherence time, the latter being inversely proportional to the spectral bandwidth. For blackbody type sources, a time resolution of about $10^{-14}$ s is necessary \cite{Mehta}. Such measurements have only been achieved with two-photon absorption techniques in semiconductors \cite{boitier}, and in ghost imaging experiments \cite{ghost}. Those are incompatible with astrophysics applications in terms of light intensity requirements. Recently, temporal photon bunching in blackbody radiation has been demonstrated with intensity correlation measurement via strong spectral filtering of the light source \cite{Kurtsiefer}. In this article, we demonstrate temporal intensity correlation measurements with light scattered by a hot (room temperature and higher) atomic vapor as a broadband pseudo-thermal source, which represents a significant step towards intensity correlation spectroscopy in astrophysics. We also show how this technique can be efficiently exploited to measure different properties of the scattered light. Notably, we infer the ratio between single and multiple scattering as a function of the optical thickness. We discuss also how the time resolution as well as the complexity of the fluorescence spectrum impact the intensity correlation measurement. Our experimental data is compared with numerical simulations, in particular to compute the single to multiple scattering ratio and the evolution of the spectrum of the scattered light.

\section{Experiment}

We study experimentally the temporal intensity correlation of light scattered by a hot rubidium vapor. For a stationary process, the temporal intensity correlation function reads
\begin{equation}
g^{(2)} (\tau) = \frac{\left\langle I(t) I(t+\tau)\right\rangle}{\left\langle I(t) \right\rangle^2}.
\end{equation}
From Cauchy-Schwarz inequalities, one can show that the $g^{(2)} (\tau)$ function is always smaller than its value at zero delay ($1<g^{(2)} (\tau)<g^{(2)} (0)$).
For chaotic light, the second order intensity correlation function can be given by the Siegert equation, 
\begin{equation}
\label{Sieg}
g^{(2)} (\tau) = 1 + \beta |g^{(1)} (\tau)|^2, 
\end{equation}
where the first order correlation $g^{(1)} (\tau)$ is the Fourier transform of the light spectrum. The factor $\beta$ is linked to the number $N$ of detected optical modes ($\beta=1/N$) and denotes the spatial coherence. For a detector radius smaller than the coherence length of the scattered light, we have $\beta = 1$. In this case, and for chaotic light, the intensity correlation at zero delay is $g^{(2)} (0) = 2$ (as $g^{(1)} (0) = 1$ by definition).
\begin{figure}[h!]
\centering
\includegraphics[width=0.45\textwidth]{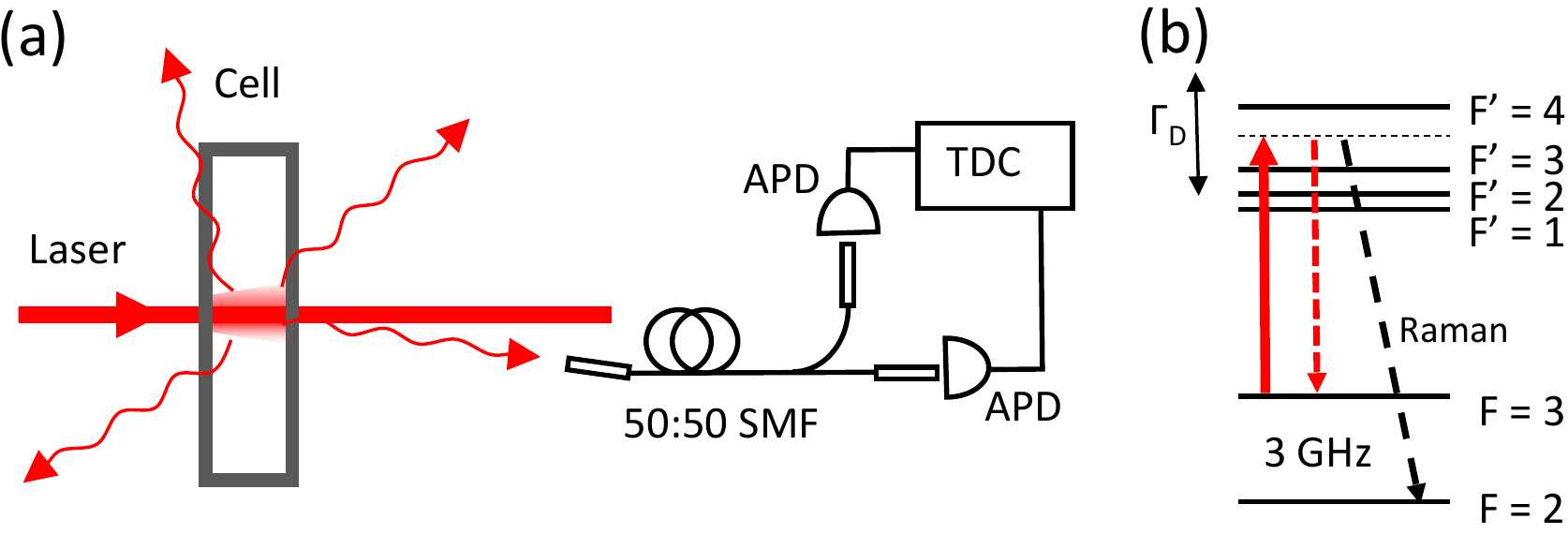}
\caption{\captionstyle
(a) Experimental setup. Light scattered in the rubidium cell is collected using a bare single mode fiber (SMF) and split with a 50/50 fiber beam splitter (BS). The two outputs of the BS are coupled to two APDs. The single counts in each APD are digitized with a TDC and analyzed with a computer. (b) Schematic of the rubidium 85 $D_2$ transition hyperfine structure. Red arrows: excitation and emission at the $\left\{F=3 \rightarrow F'=3,\  F=3 \rightarrow F'=4 \right\}$ crossover frequency. Black dashed arrow: emission through Raman scattering. $\Gamma_D$ gives the magnitude of the Doppler broadening.}
\label{fig:schema}
\end{figure}

The schematics of our experiment is depicted in Fig.~\ref{fig:schema}(a). We use a rubidium cell containing a natural mixture of the two isotopes $^{85}$Rb and $^{87}$Rb \cite{Steck}, having a radius of 5 cm and a thickness of 5 mm. The beam of a commercial external cavity diode laser (Toptica DL Pro) is passed through the center of the cell and has a waist of 2 mm and a power of 900 $\mu W$. In the center of the laser beam, the intensity is 14.3 mW/cm$^2$, which is about 9 times larger than the saturation intensity but due to the large Doppler broadening ($\approx$ 250MHz), only a small fraction of atoms are resonantly excited and we therefore neglect the inelastic scattering corresponding to the Mollow triplet \cite{Jurczak}. The scattered light is collected with a single mode fiber (SMF), placed at a distance $L= 25$ cm after the cell, with an angle $\theta=5.6^{\circ}$ from the laser propagation direction. No collimation lens is used, and the collection of the scattered light is thus determined by the fiber core. The distance is chosen such that the conditions for maximum spatial coherence ($\beta=1$) are satisfied \cite{Nakayama:10}. Indeed, the correlation length is estimated to be $l_c=\lambda L/(\pi s)= 62 \:\mu m $ for a source radius $s = 1 $ mm, which is much larger than the SMF mode-field diameter ($\approx 5-6 \:\mu m$). The coupled light is then split at a 50/50 fiber beam splitter and detected using two single-photon avalanche photo diodes (APD, SPCM-AQRH from Excelitas Technologies \cite{APD}). Those APDs feature a quantum efficiency of about 60$\%$ at 780 nm. To build up the $g^{(2)} (\tau)$ measurements, time tags, with a resolution of 160 ps,  are obtained from a multichannel time-to-digital converter (TDC, ID800 from IdQuantique), and sent for analysis to a computer. For the measurements presented here, the laser frequency is locked at the $\left\{F=3 \rightarrow F'=3,\  F=3 \rightarrow F'=4 \right\}$ crossover frequency of the rubidium 85 $D_2$ line. The rubidium cell is placed in an oven in order to vary the saturation vapor pressure and thus the atomic density. The temperature is varied from $20^{\circ}$C to $78^{\circ}$C, yielding an optical thickness $b$ in the range $0.07< b <12$. The detectors count rates range from $2.7 \times 10^{3}$ s$^{-1}$  and $4 \times 10^{4}$ s$^{-1}$, well above the dark count rate ($<$ 100 s$^{-1}$). In order to keep the signal high compared to stray light ($\approx 2000$ cts/s) and detector dark counts, no polarizer is used after the rubidium cell. The integration duration, ranging from 10 hours to a couple of days, has been adapted to observe a total number of counts of at least $5 \times 10^{8}$ for every series of measurements. The temperature of the cell has been obtained by fitting the transmission through the cell as a function of the laser frequency, with low power \cite{0953-4075-40-1-017}, in the full range of the rubidium 85 and 87 $D_2$ lines. The temperature-dependent atomic density is therefore inferred from the optical opacity. 

\section{Results}

We show in Fig.~\ref{fig:g2}(a) an example of the intensity correlation measured for $b=0.38$. At first, we clearly see the modification of the photon statistics induced by the scattering of the light by the atoms with strong bunching at zero delay. We also see that the ideal value $g^{(2)} (0) = 2$ is not reached and that the correlation decay cannot be described with a simple Gaussian function as could be expected from the Doppler frequency broadening \cite{PhysRevA.53.3469} (collisional broadening is negligible for our sample). In the following, the shape of the measured intensity correlation and the evolution of the temporal contrast defined as $g^{(2)} (0)-1$, will be discussed separately in sections \ref{sub_mult} and \ref{sub_cont}.

\subsection{Single to multiple scattering ratio}
\label{sub_mult}
\begin{figure*}
\centering
\includegraphics[width=0.9\textwidth]{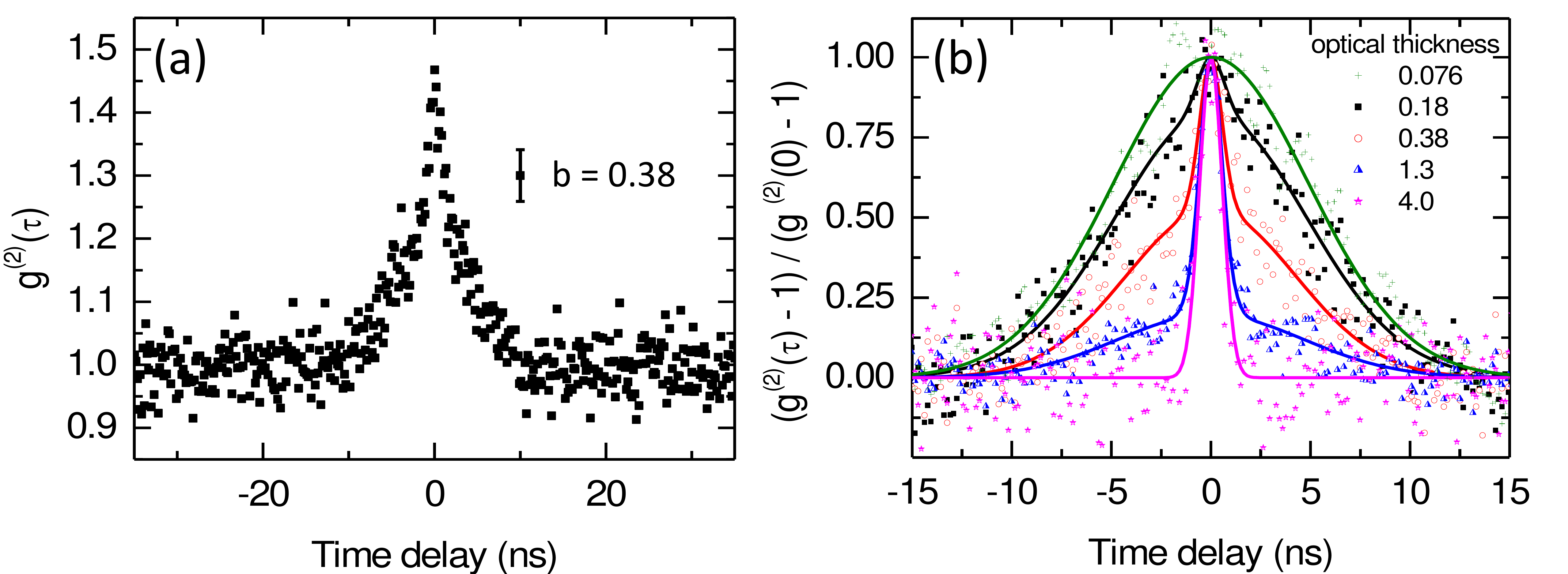}
\caption{\captionstyle
(a) Measurement of the second order intensity correlation $g^{(2)} (\tau) $ as a function of the time delay. (b) Experimental data normalized by the contrast (data points). Numerical fit according to Eq. \ref{g2}, with the ratio $\sigma_m/\sigma_s$ fixed by the experimental measurement of the detection angle (full lines).}
\label{fig:g2}
\end{figure*}

In order to compare the measurements made at different optical thicknesses, we normalize the $g^{(2)} (\tau)-1$ curves by the temporal contrast. The resulting intensity correlations are shown in Fig.~\ref{fig:g2}(b). The measured correlations clearly reveal the presence of different correlation time scales. At large optical thickness, the scattered light measured in transmission is mainly composed of photons that have scattered more than once. The Doppler broadening of the scattered light is on average isotropic and one expects a complete frequency redistribution (CFR) \cite{Molisch}. Fitting the $g^{(2)} (\tau)$ data obtained for $b=4$ with a simple Gaussian distribution gives a coherence time $\tau_c$ = 531 ps (pink line) that we relate to a Gaussian frequency broadening  $\sigma$ = 212 MHz from $\tau_c= 1/(2 \pi \sqrt{2} \sigma)$. This value is close to the Doppler width expected in the CFR regime at $64^{\circ}$C, $\sigma_m = \sqrt{\frac{k_B T}{m c^2}} \nu_0 = 230$ MHz, with $\nu_0$ the atomic transition frequency, $k_B$ the Boltzman constant, $T$ the temperature, $m$ the mass of the rubidium atom and $c$ the speed of light. 

By contrast, for a small optical thickness, most of the detected photons have only scattered once. There, the Doppler broadening is strongly dependent on the scattering angle. Contrary to the cold atom case, the standard deviation of the longitudinal velocity distribution of absorbing atoms is given by the natural linewidth \cite{Martine2015}, much smaller than the Doppler width, such that only the transverse velocity distribution contributes significantly to the Doppler broadening. Taking this into account, we expect the Doppler broadening, after single scattering, to be $\sigma_s= \sin{\theta} \: \sigma_m$. This expression implies that we can neglect the spectral broadening through Doppler effect for the light scattered in the laser propagation direction ($\theta= 0$) and, less intuitively, in the opposite direction ($\theta= \pi$) (back scattering) \footnote{No Doppler broadening does not mean no Doppler effect. Indeed the largest Doppler frequency shift applies to the back-scattered light.}. In the following, we make use of this strong angle dependence of the Doppler broadening, by detecting the scattered light with a small angle ($\theta= 5.6^{\circ}$) with respect to the laser propagation direction, in order to achieve a quantitative measurement of the multiple scattering ratio. Indeed, in the single scattering regime we expect the Doppler broadening to be about an order of magnitude smaller than in the multiple scattering regime. This is confirmed by the fit with a single Gaussian of the $g^{(2)} (\tau)$ data for the smallest optical thickness $b=0.076$, which infers a broadening of 23 MHz. This is close to the theoretical value $\sigma_s = 21$ MHz for $T=20^{\circ}$C.   

\begin{figure}[t]
\centering
\includegraphics[width=\width\textwidth]{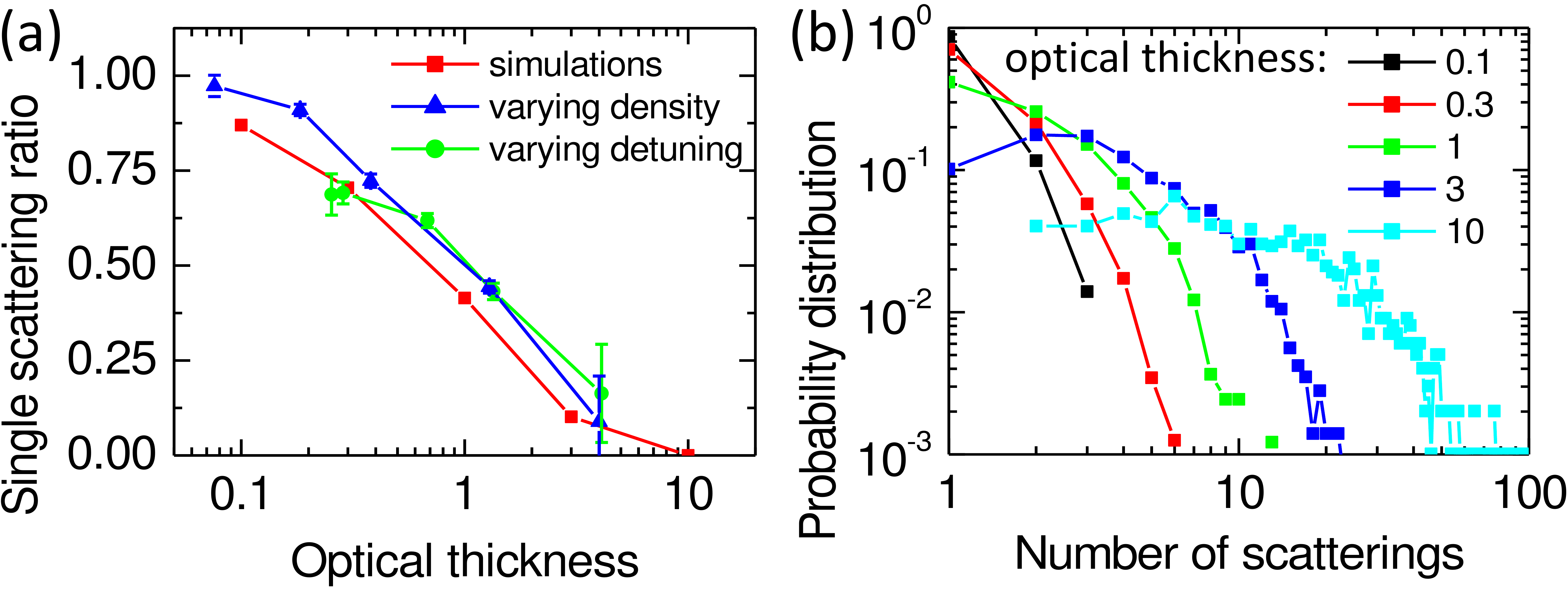}
\caption{\captionstyle
(a) Single scattering ratio $R_s$ as a function of the optical thickness obtained from random walk simulations (red squares), and experimentally by varying the vapor atomic density (blue triangles) or the laser detuning (green dots). (b) Simulated probability that a photon detected in the solid angle $0^{\circ}<\theta<5^{\circ}$ has scattered $n$ times, as a function of the optical thickness.}
\label{fig:ratio}
\end{figure}

In the intermediate regime between single and multiple scattering, we therefore expect to observe a Doppler-broadened spectra with two easily distinguishable components from multiple and single scattering,
\begin{equation}
\label{Pw}
P(\nu)   \propto  a \exp\left(- \frac{(\nu - \nu_0)^2}{2  \sigma_m^2}\right) + b \exp\left(- \frac{(\nu - \nu_0)^2}{2  \sigma_s^2}\right),
\end{equation}
with $a$ and $b$ the relative amplitudes of the two components ($a+b = 1$) and $\nu_0$ the central frequency of the fluorescence spectrum. With this definition, the relative weights of multiple and single scattering (proportional to the area of each Gaussian) are $R_m = a \sigma_m /(a \sigma_m + b \sigma_s)$ and $R_s = b \sigma_s /(a \sigma_m + b \sigma_s)$. The Fourier transform of Eq.~\ref{Pw} gives $g^{(1)} (\tau)$, then we obtain an expression for the temporal intensity correlation from Eq.~\ref{Sieg},
\begin{multline}
\label{g2}
g^{(2)} (\tau) - 1 =  R_m^2 e^{- \frac{\tau^2}{2\tau_m^2}} + R_s^2 e^{- \frac{\tau^2}{2\tau_s^2}} + 2 R_m R_s e^{- \frac{\tau^2}{4 \tau_m \tau_s}},
\end{multline}
with $\tau_{m,s}= 1/(2 \pi \sqrt{2} \sigma_{m,s})$ the coherence times in the multiple and single scattering regimes. Note that, as $g^{(2)} (\tau)$ is a quadratic function of $g^{(1)} (\tau)$, the two components of the optical spectrum give rise to the superposition of three decay times. Fig.~\ref{fig:g2}(b) shows the fit of the measured $g^{(2)} (\tau)$ by Eq.~\ref{g2} (full lines), with all parameters set free, apart from the ratio $\sigma_s / \sigma_m = \sin{\theta}$, fixed according to the experimental detection angle. The calculated fits as well as the experimental data have been normalized by the contrast ($R_m^2+R_s^2+R_mR_s$). On Fig.~\ref{fig:ratio}(a), we show the deduced single scattering ratio $R_s$ as a function of the optical thickness. 

We also performed measurements for which we varied the optical thickness by changing the laser frequency instead of the temperature, with detunings smaller than the Doppler width. Those measurements gave similar values of  $R_s$ and $R_m$ as a function of the optical thickness [see Fig.~\ref{fig:ratio}(a)].

In order to determine the accuracy of this measurement, we performed random walk simulations with a Monte Carlo routine based on first principles \cite{Martine2015}. The simulations use a two-level atom model and a Voigt absorption profile. Indeed, the probability $P(v_{\parallel})$ that an atom, with a velocity $v_{\parallel}$ along the photon propagation direction, absorbs an incoming photon, is given by the product of the Doppler-shifted Lorentzian atomic lineshape, and the Maxwell-Boltzmann distribution of $v_{\parallel}$,
\begin{equation}
\label{Voigt}
\begin{aligned}
P(v_{\parallel}) \propto  \frac{1}{1+4 \frac{\delta^2}{\Gamma^2}} e^{-v_{\parallel}^2/u^2},
\end{aligned}
\end{equation}
with $\delta$ the detuning of the incoming photon in the atomic rest frame, $\Gamma$ the natural linewidth, and $u = \sqrt{2 k_B T/m}$ the half-width of the Maxwell-Boltzmann velocity distribution. A cell with the same dimensions of the cell of the experiment is considered (radius 5 cm, thickness 5mm). The incident beam is infinitely narrow and spectraly monochromatic. By computing the path and the Doppler frequency shift of a large number of photons, we are able to build up statistics about the number of scattering events and frequency distribution of the output photons. In Fig.~\ref{fig:ratio}(b) we show, for different values of optical thicknesses, the probability that a photon, initially at resonance, and detected in the solid angle $0^{\circ}<\theta<5^{\circ}$, has scattered $n$ times. The solid angle is chosen larger than in the experiment in order to reduce the computing time. The data shown here is not very sensitive to this angle, especially at small optical thickness. The deduced values of $R_s$, the single scattering ratio ($n = 1$), are reported in Fig.~\ref{fig:ratio}(a) (red squares). We can see that the ratio extracted from our experimental measurement is well reproduced by the simulations.

\subsection{Contrast of the temporal correlation}
\label{sub_cont}

The previous simulations do not take into account the multilevel structure of the excited states of the rubidium and the spontaneous Raman scattering between the two hyperfine ground states ($F=3 \rightarrow F=2$) [see Fig.~\ref{fig:schema}(b)], which plays an important role in the decrease of the contrast, as explained in the following. 

\begin{figure}[t]
\centering
\includegraphics[width=0.4\textwidth]{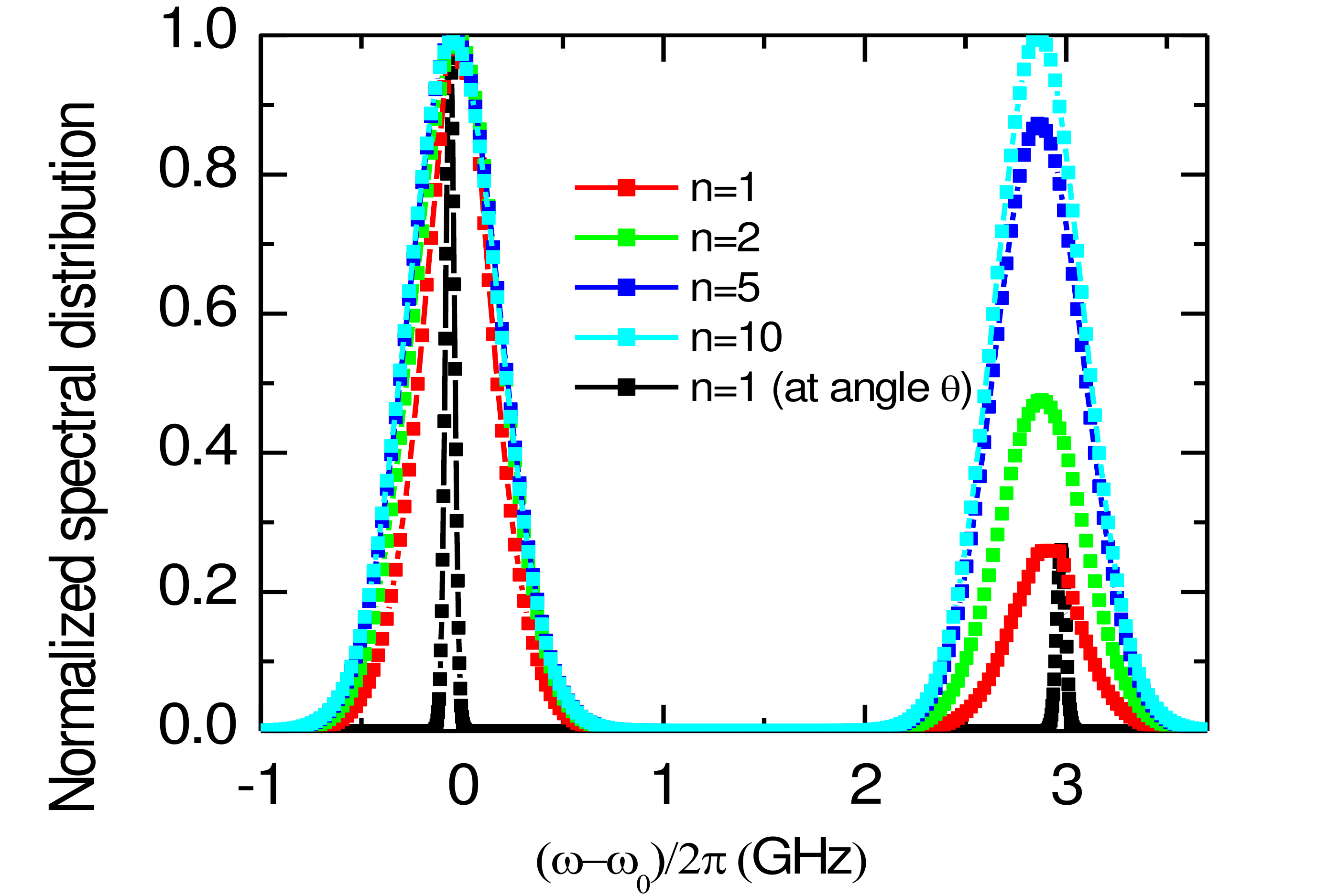}
\caption{\captionstyle
Calculated evolution of the emission spectrum for scattering events ranging from $n$ = 1 to $n$ = 10, averaged in all directions at room temperature, by taking into account the multilevel structure of rubidium and the Raman scattering. The peak on the left corresponds to emission from Rayleigh scattering, the peak on the right correspond to emission from anti-Stokes Raman scattering. Black squares: Spectrum after the first scattering for $\theta = 5.6^{\circ}$}. 
\label{fig:spectre}
\end{figure}

In Fig.~\ref{fig:spectre}, we show the evolution of the emission spectrum of rubidium at room temperature, that we calculate numerically, for an infinite medium, by taking into account the multilevel structure and the Raman scattering. The spectra are obtained by calculating the convolution between the laser profile, the frequency dependent atomic cross-section, the Maxwell-Boltzman distribution of the atomic velocities, as well as the different transition factors within the multilevel structure. The numerical method used for single and multiple scattering is detailed in \cite{mercadier2013} [see Eqs. (19) and (21)]. We used, as in the experiment, a 1 MHz broad excitation at the rubidium 85 crossover frequency. For the spectrum after the first scattering, we consider two cases. First, an emission at $\theta = 90^{\circ}$, which gives a good estimation of the spectrum averaged in all the scattering angles (see red curve in Fig.~\ref{fig:spectre}). This is necessary to calculate the next spectra \cite{mercadier2013} but does not reflect the spectrum measured in our experiment. In order to simulate the single-scattering spectrum as measured in the experiment at low optical thickness, we include the angle dependence by considering the energy conservation of a photon scattered at $\theta = 5.6^{\circ}$ from the incident laser propagation direction (see black curve in Fig.~\ref{fig:spectre}) \footnote{From equation 19 in \cite{mercadier2013}, we replace $v_x$ and $v_y$ by $\sin{\theta} v_x$ and $(1-\cos{\theta}) v_y$ in the Dirac distribution that denotes the energy conservation.}. For the multiple scattering regime, we calculate the evolution of the spectra after up to 10 scattering events, which was enough to obtain a complete frequency redistribution (no noticeable change of Raman scattering ratio, neither change of the spectrum linewidth was observed beyond).

\begin{figure}[h!]
\centering
\includegraphics[width=\width\textwidth]{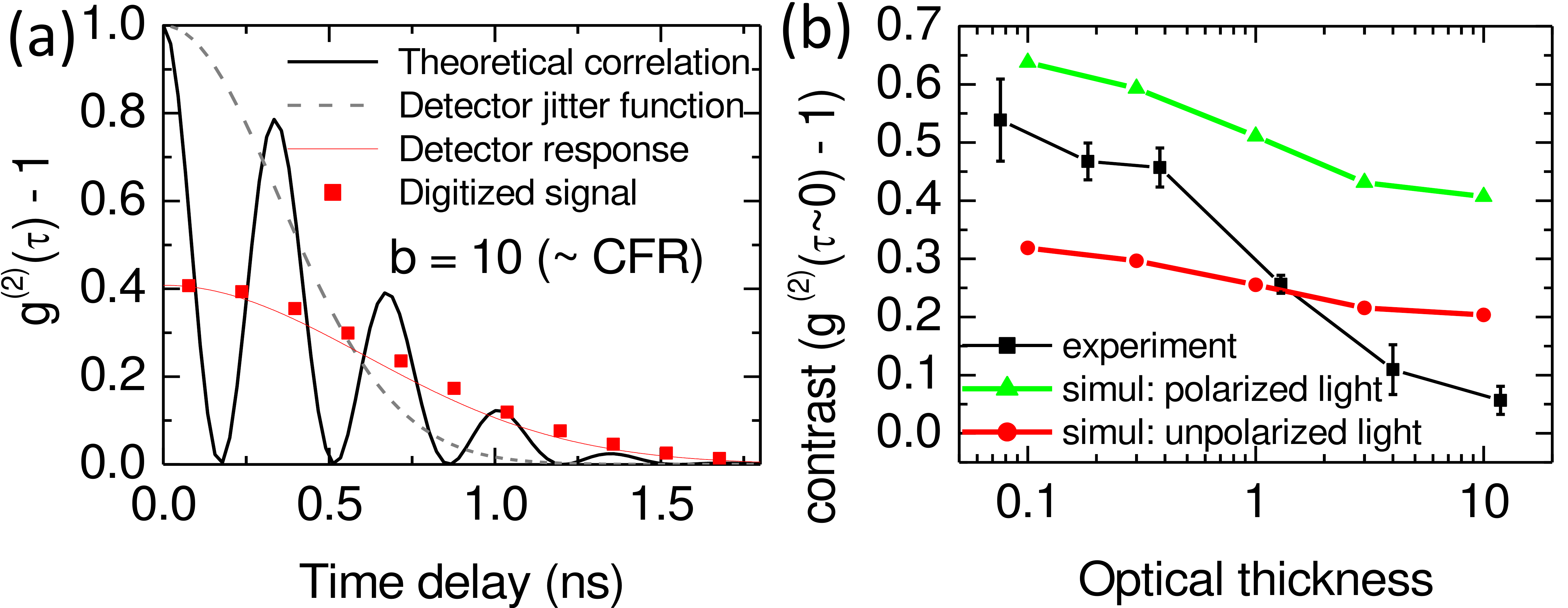}
\caption{\captionstyle
(a) Calculated $g^{(2)} (\tau)$ curve for $b=10$ (black line), 350 ps detector jitter function (grey dashed line), convolution between $g^{(2)} (\tau)$ and the detector jitter (red line), simulation of the signal as given by the time to digital converter (red squares). (b) Contrast of the experimental measurement (black squares) and simulated contrast for a fully polarized light (green triangles) and unpolarized light (red dots).}
\label{fig:contrast}
\end{figure}

In order to determine the impact of the multilevel structure of the rubidium and the Raman scattering on our intensity correlation measurement, we estimate the spectrum of the detected light as a function of the optical thickness by multiplying the spectra shown in Fig.~\ref{fig:spectre} with their respective weights as given in Fig. \ref{fig:ratio}(b). As an example, Fig.~\ref{fig:contrast}(a) shows the theoretical correlation function for $b=10$ deduced from Eq.~\ref{Sieg}.

By itself, the impact of the hyperfine structure of the \textit{excited} states only gives rise to a slightly larger effective broadening of the spectrum. This induces a negligible decrease of the correlation decay time. 
The hyperfine structure of the ground state, on the contrary, has a dramatic impact. We calculate that  $26 \%$ of the photons are scattered through Raman scattering at the first scattering event. This ratio increases with the number of scattering events (up to $ \approx 50 \%$). One of the consequences is a small decrease of the overall atomic cross section $\sigma_{sc}$ that is not included in the random walk simulations.

The main impact of the Raman scattering on our measurement lies in the beating at 3 GHz between the Rayleigh and the Raman components of the fluorescence spectrum. This results in an oscillatory behavior of the intensity correlation, on a timescale that is not accessible with our experimental setup. In Fig.~\ref{fig:contrast}(a) we show, as an example, the convolution between $g^{(2)} (\tau)$, for $b=10$, and a Gaussian function of FWMH = 350 ps (grey dashed line) that simulates our detector timming jitter (red line) \cite{APD}. The red squares simulate the time binning (160 ps) imposed by the TDC. If the timing resolution suppresses the oscillations, it also induces a reduction of the measured contrast, proportional to the Raman scattering rate. Fig.~\ref{fig:contrast}(b) shows the resulting contrast for the calculated $g^{(2)} (\tau)$  (green triangles) and the contrast of the measured $g^{(2)} (\tau)$ (black squares). The calculated contrast has also been divided by a factor two (red dots) to show the expected values with unpolarized light \cite{Brown300} (no polarizer was placed between the cell and the detector during the experiment). Note that additional measurements with faster detectors (70 ps jitter) but lower quantum efficiency revealed the oscillation of $g^{(2)} (\tau)$ due to Raman scattering (not shown). However, the poor signal-to-noise ratio did not allow us to make any quantitative measurement of the Raman scattering rate.

Except for the lowest optical thickness $b=0.07$, for which the contrast is low because of a high relative amount of stray light ($\approx 40 \%$), the high contrast at low $b$ is in agreement with the expectation for partially polarized light. The decrease with $b$ is consistent with a decrease of the degree of polarization due to multiple scattering. However, at high $b$, the contrast decreases to values lower than expected.  We have checked that the stray light (e.g., from the laser diode amplified spontaneous emission), or the level of fluorescence, as well as the increase of the source size with $b$ due to diffusion \cite{PhysRevE.90.052114}, are not responsible for the anomalous decrease of contrast. It should be mentioned that the same phenomena has been predicted \cite{PhysRevA.68.013411} and observed in a cold atom experiment \cite{Stites:04} where radiation trapping (with optical thicknesses as low as 0.4) was considered as the origin of the $g^{(2)} (0)$ decay. Nevertheless, this assumption was contradicted in a more recent publication, where full contrast intensity correlation was reported in optical molasses with an optical thickness going up to 3 \cite{Nakayama:10}. Further measurements with a polarizer and a larger integration time would allow us to discard polarization effects, in order to get a better understanding of this phenomena. We could also reproduce the experiment with even lower laser intensities to definitely suppress inelastic scattering \cite{PhysRev.188.1969} or other nonlinear effects. Ultra narrow spectral filtering could also be used to isolate the Rayleigh component of the spectrum and cancel the $g^2(\tau)$ oscillations caused by Raman scattering. 

\begin{figure}[t]
\centering
\includegraphics[width=0.4\textwidth]{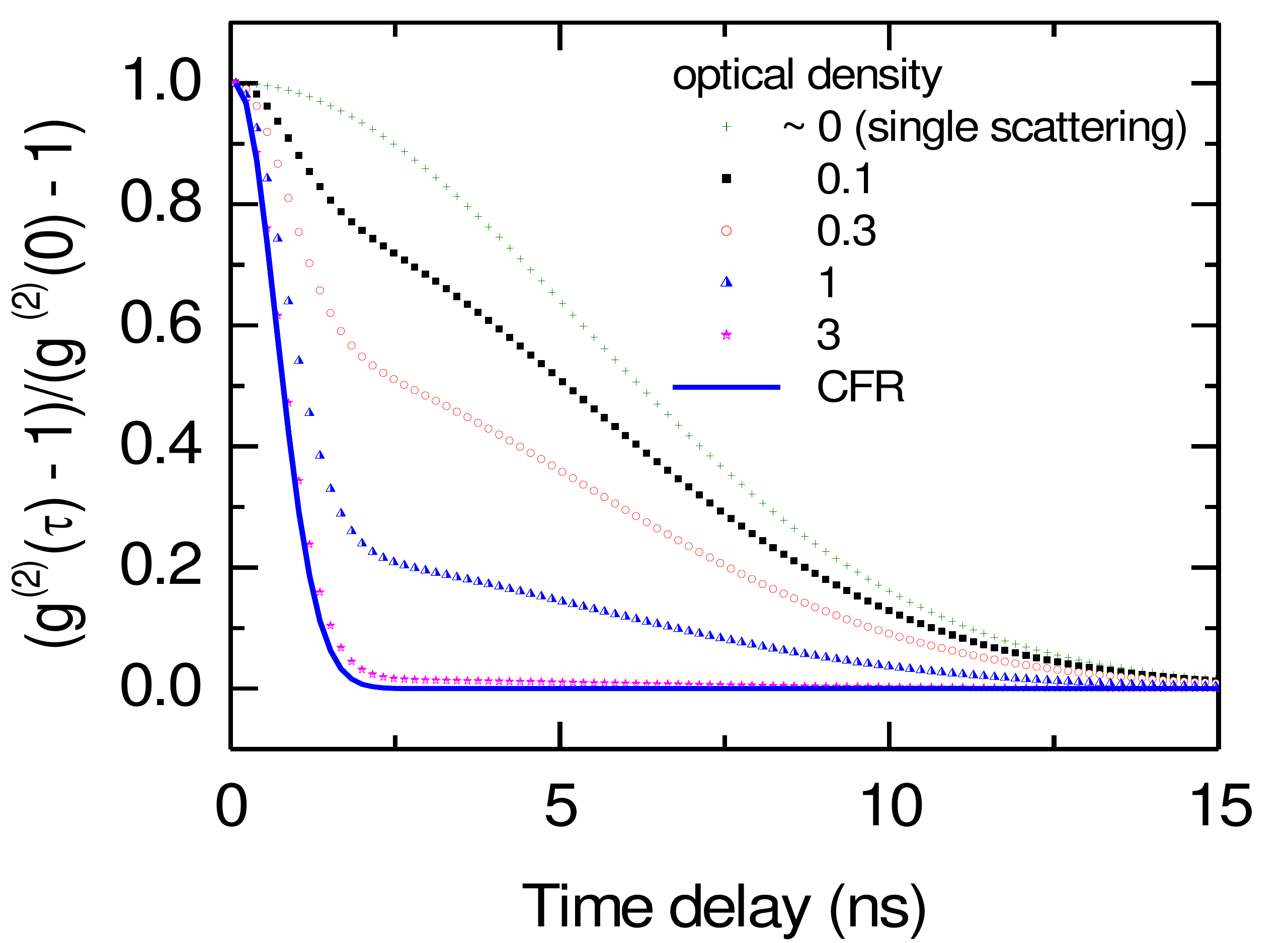}
\caption{\captionstyle
 Simulation of the second order intensity correlation $g^{(2)} (\tau) - 1$ as a function of the time delay for different optical thicknesses, by taking the detector jitter and time resolution into account, and calculated from the spectra shown in Fig.~\ref{fig:spectre} multiplied by their respective weights as given in Fig.~\ref{fig:ratio}(b). The data has been normalized by the contrast shown in Fig.~\ref{fig:contrast}(b).}
\label{fig:g2_simu}
\end{figure}

To summarize our simulations of the single to multiple scattering ratio and our calculation of the scattered light spectrum, we show in Fig.~\ref{fig:g2_simu} the normalized $g^{(2)} (\tau)$ curves calculated for the different values of optical thickness, including the multilevel structure, the Raman scattering, and the detection time resolution. The single scattering and CFR regimes are respectively calculated from the spectra after $n=1$ and $n=10$ scattering events. Despite the presence of non-elastic scattering and the complex multilevel structure of rubidium, we can obtain a very good agreement between the simulated and measured $g^{(2)} (\tau)$ curves [Fig. \ref{fig:g2}(b)].

\section{Conclusion}

Intensity correlation measurements with broadband light is challenging as it requires a detection scheme featuring a high timing resolution. Here, we have demonstrated intensity correlation with a hot atomic vapor with a coherence time on the order of tenths of nanoseconds, i.e., much lower than with cold atoms. The reported results can find important repercussions in applications related to astrophysics, where sub-GHz spectral filtering is still challenging in the visible range \cite{Kurtsiefer}, and where spectral features such as astrophysical lasers may be investigated \cite{Johansson2005361,1301176}. We have shown that we were able to quantitatively measure the single to multiple scattering ratio and demonstrated a good agreement between experimental and simulated results. These measurements may also be useful, in further studies, to investigate the polarization or the anomalous correlation of the scattered light at different optical thicknesses. In particular, it is still not clear why the contrast is reduced at high optical thicknesses. One may question the impact of the non-Gaussian statistics of the photon step length in atomic vapors \cite{Levy_flight}. Eventually, there might also be a subtle change of the correlation function due to the Zeeman degeneracy \cite{PhysRevLett.98.083601,PhysRevA.92.013819,PhysRevA.92.033853}.

\begin{acknowledgments}
We thank A. Eloy and A. Siciak for help with the data acquisition, and M. Lintz and P. Nu\~nez for many useful discussions. A.D. is supported by Campus France (Program n.CF-PRESTIGE-18-2015). We also thank the Doeblin Federation for financial support. T.P.S. thanks the support of Brazilian agency CNPq (Conselho Nacional de Desenvolvimento Cient\'ifico e Tecnol\'ogico).
\end{acknowledgments}

\bibliographystyle{C:/Users/Robin/Documents/rapports-presentations/hot_rb_G2_paper_first_part/apsrev4-1}
\bibliography{C:/Users/Robin/Documents/rapports-presentations/hot_rb_G2_paper_first_part/bibi}

\end{document}